\documentclass[pra,aps,showpacs,epsf,twocolumn,superscriptaddress]{revtex4}

\usepackage{graphicx}
\usepackage{epsfig}
\usepackage{amsmath}
\usepackage{amssymb}

\begin{document}
\title{Collisional properties of weakly bound heteronuclear dimers}
\author{B.~Marcelis}
\affiliation{Eindhoven University of Technology, P.O.~Box~513, 5600~MB  Eindhoven, The Netherlands}
\affiliation{\mbox{Laboratoire Physique Th\'eorique et Mod\`eles Statistique, Universit\'e Paris Sud, CNRS, 91405~Orsay, France}}
\author{S.J.J.M.F.~Kokkelmans}
\affiliation{Eindhoven University of Technology, P.O.~Box~513, 5600~MB  Eindhoven, The Netherlands}
\author{G.V.~Shlyapnikov}
\affiliation{\mbox{Laboratoire Physique Th\'eorique et Mod\`eles Statistique, Universit\'e Paris Sud, CNRS, 91405~Orsay, France}}
\affiliation{\mbox {Van der Waals-Zeeman Institute,University of Amsterdam,
Valckenierstraat 65/67, 1018 XE Amsterdam, The Netherlands}}
\author{D.S.~Petrov}
\affiliation{\mbox{Laboratoire Physique Th\'eorique et Mod\`eles Statistique, Universit\'e Paris Sud, CNRS, 91405~Orsay, France}}
\affiliation{Russian Research Center Kurchatov Institute, Kurchatov Square, 123182 Moscow, Russia}
\date{\today}
\hyphenation{Fesh-bach}

\begin{abstract}
We consider collisional properties of weakly bound heteronuclear molecules (dimers) formed in a two-species mixture of atoms 
with a large mass difference. We focus on dimers containing light fermionic atoms as they manifest
collisional stability due to an effective dimer-dimer repulsion originating from the exchange of the light atoms. 
In order to solve the dimer-dimer scattering problem we develop a theoretical approach, 
which provides a physically transparent and quantitative description of this four-atom system in terms of three- and two-body 
observables. We calculate the elastic scattering amplitude and the rates of inelastic processes such as the trimer formation and 
the relaxation of dimers into deeply bound molecular states. Irrespective of whether the heavy atoms are bosons or fermions, the inelastic 
rate can be significantly lower than the rate of elastic collisions. Moreover, the measurement of the inelastic rate which is a four-body 
observable, can be an efficient and precise tool for determining three-body observables such as the three-body parameter, 
positions of Efimov states, and their lifetimes.
\end{abstract}


\maketitle

\section{Introduction}
For many years the production of ultracold molecules was a long term goal of various experiments \cite{doyle04,dulieu06}. 
A breakthrough in this field has recently been achieved by combining the ability to cool ultracold atomic gases to very 
low temperatures and the ability to associate pairs of them into weakly bound dimers by using Feshbach resonances 
\cite{donley02,herbig03,xu03,greiner03,jochim03,zwierlein03,bourdel04,durr04,ospelkaus06}. This approach turns out to be 
efficient for obtaining homonuclear weakly bound molecules out of both fermionic and bosonic atoms. In the former case 
the Pauli exclusion principle for the constituent atoms makes the molecular gas remarkably stable \cite{petrov04,petrov05,petrov_v2_05}, 
which has allowed the observation of Bose-Einstein condensation (BEC) of molecules and investigation of their superfluid 
properties \cite{greiner03,jochim03,zwierlein03,bourdel04}. Molecules consisting of bosonic atoms are sufficiently stable 
only if they are very dilute or completely isolated from each other, for example, by an optical lattice. However, 
exploration of such molecules, and strongly interacting bosonic atoms in general, is still on the short list of many 
experiments motivated by the potential opportunity to study the Efimov few-body physics \cite{efimov70,jensen04,braaten07}.
The evidence for the existence of Efimov states has recently been obtained in experiments with cesium atoms in Innsbruck
\cite{kraemer06,naegerl06}.

The tunability of the interspecies interaction in heteronuclear mixtures has been demonstrated in a number of experiments 
\cite{inouye04,stan04,salomon04,inguscio06,inguscio06_v2,ospelkaus06,ospelkaus_v2_06,grimm07}, which is an important step towards obtaining 
ultracold heteronuclear molecules. This is believed to be a rewarding project owing to a permanent electric dipole moment that 
they posses in their ground state. There exists a wide range of proposals to use such molecules for studies of novel quantum 
phases, realizing quantum computing schemes, testing fundamental physical concepts, etc. (see, for example, 
Refs. \cite{doyle04,dulieu06,ospelkaus06}, and references therein).

Taking into account the success in associating homonuclear molecules from ultracold atomic samples, it seems reasonable 
to generalize this technique to mixtures of different atoms. So far, this scheme has been used to create weakly bound K-Rb 
molecules in an optical lattice \cite{ospelkaus06}. The association of atoms into a weakly bound molecular state can be 
applied, in principle, in many other experiments studying ultracold mixtures with various fermionic and bosonic isotopes. 
The main concern of this scheme is the collisional stability of the molecules that are associated into highly excited 
rovibrational states. Stabilizing such weakly bound dimers without isolating them from each other opens up a number of 
interesting research directions such as the study of novel quantum phases in bosonic molecular gases \cite{PAPSS} and in the presently 
completely unexplored fermionic molecular gases. It would also shed light on many intriguing questions related to the few-body 
physics with atoms of different masses and statistics.

The theoretical description of elastic and inelastic dimer-dimer collisions requires the solution of a four-atom problem which, 
in a general case, can hardly be accomplished. An efficient theoretical approach based on the short-range character of  
the interatomic forces allows the calculation of scattering properties of weakly bound homonuclear and heteronuclear bosonic 
molecules consisting of fermionic atoms \cite{petrov04,petrov05,petrov_v2_05}. However, this approach is limited to the case where the mass ratio of constituent atoms is smaller
than a critical value of 13.6 and the subsystem of one light and two heavy fermions does not exhibit the Efimov effect. Then the scattering amplitude is a universal function of
the atom-atom scattering length $a$ and the atom masses, and the so-called three-body 
parameter is not needed. For larger mass ratios or in the case of bosonic constituent atoms one has to take into account the 
existence of Efimov three-body bound states, the three-body parameter comes into play, and it is still debated whether an additional 
four-body parameter is required to describe four-body observables \cite{platter04,yamashita06}.

In this paper we discuss the scattering properties of weakly bound dimers which consist of atoms with a large mass difference. 
Both the light and heavy atoms are in a single quantum state, and the weakly bound molecular states are formed in their gaseous mixture 
at a large and positive scattering length for the light-heavy interaction.  Our attention is focused on dimers containing light 
fermionic atoms as these dimers should be collisionally stable due to an effective dimer-dimer repulsion originating from the exchange of the light atoms.
We develop an approach for solving the dimer-dimer scattering problem, which in addition to the short-range character 
of the interatomic interactions, relies on a large difference in the masses of atoms forming a dimer. The latter circumstance 
allows us to use the Born-Oppenheimer adiabatic approximation and integrate out the motion of one light atom substituting it 
by an effective potential acting on the heavy atoms. This reduces the dimer-dimer scattering to a three-body problem, which can be solved by 
developing the method presented in Ref.~\cite{petrov03}.

Such a hybrid Born-Oppenheimer (HBO) approach provides a transparent description of the four-body system of two light and two 
heavy atoms. It allows one to express four-body observables in terms of the light-heavy scattering length $a$ and the three-body 
parameter for the subsystem of one light and two heavy atoms. Remarkably, this relation can be inverted and can be used to retrieve  
information on the three-body system from the collisional properties of the dimers. As we show in this paper, the lifetime of 
a gas of dimers can significantly exceed the lifetime of an atomic mixture or an atom-dimer mixture. This suggests that indirect
measurements of three-body effects through four-body observables can be more efficient and precise than measurements using 
three-body systems directly.

We present a detailed analysis of elastic and inelastic dimer-dimer collisions, based on the HBO approximation. 
We calculate the $s$-wave ($p$-wave) amplitudes of elastic scattering between ultracold bosonic (fermionic) dimers 
and discuss the relaxation of dimers into deep bound states as well as the formation of trimer bound states in dimer-dimer 
collisions. As shown, the inelastic rates are significantly suppressed by the long-range exchange repulsion between the dimers.

The paper is organized as follows. In Sec.~\ref{II} we introduce the Born-Oppenheimer approximation for three- and 
four-body problems in which both light atoms are adiabatically eliminated and in Sec.~\ref{III} we discuss possible inelastic decay mechanisms in 
dimer-dimer collisions. We then justify the need for a more sophisticated HBO approach to describe the dimer-dimer scattering. 
Section \ref{IV} contains the derivation of the HBO equation for the four-body problem and in Sec.~\ref{V} we show how it 
can be used for calculating elastic and inelastic dimer-dimer collisional properties. We present the results 
for the dimer-dimer scattering amplitude as a function of the atomic scattering length and the three-body parameter in the case 
of bosonic and fermionic molecules. We also discuss implications of the results for experimental observation of dimers, Efimov 
trimers, and related few-body physics. In Sec.~\ref{VI} we discuss the validity of the HBO approach and compare the HBO results 
with exact calculations for moderate mass ratios and in Sec.~\ref{VII} we conclude.

\section{Born-Oppenheimer approximation for three- and four-body systems}\label{II}

Let us consider the interaction between two weakly bound dimers consisting of atoms with very different masses and discuss 
the dimer-dimer scattering properties. A natural way to benefit from the large mass ratio is to use the Born-Oppenheimer 
(BO) approximation \cite{born27}, where one assumes that the state of fast light atoms adiabatically adjusts itself to the 
positions of the slow heavy atoms. In Ref.~\cite{fonseca79} the BO approach has been developed for a three-body system with an emphasis 
on the Efimov effect. The four-body case has been briefly discussed in Ref.~\cite{petrov_v2_05}. 

Within the BO approach the four-body problem can be split into two parts. First, one calculates the wavefunctions and binding energies 
of the light atoms in the field of the heavy atoms fixed at a distance ${\bf R}$ from each other. The sum of the light-atom binding 
energies gives the potential energy surface for the heavy atoms, and the second step of the BO procedure is to consider the motion 
of the heavy atoms in this effective potential.

We assume that the interaction between the light atoms is not resonant and can be neglected. Then each of the light atoms independently 
moves in the field created by the pair of heavy atoms. For the interaction between light and heavy atoms we use the zero-range 
Bethe-Peierls approach \cite{bethe35} assuming that the motion of light atoms is free everywhere except for a vanishing distance between the light and heavy atoms. 
This approach makes sense for $R\gg \tilde R_e$ and 
\begin{equation}    \label{aRe}
a\gg R_e,
\end{equation} 
where $a$ is the scattering length for the light-heavy interaction and $R_e,\tilde R_e$ are the characteristic radii of the 
light-heavy and heavy-heavy interatomic potentials, respectively.
Then there are two bound states of a light atom in the field of two heavy ones: the gerade state 
$(+)$ with the wavefunction remaining unchanged under permutation of the heavy atoms $({\bf R}\rightarrow -{\bf R})$,
and the ungerade state $(-)$ with the wavefunction changing its sign under this operation. The corresponding wavefunctions are given by
\begin{equation}       \label{BOWaveFunctions}
\psi_{\bf R}^{\pm}({\bf r}) = {\cal N}_\pm \left( \frac{e^{- \kappa_{\pm}(R)|{\bf r}-{\bf R}/2|}}{|{\bf r}-{\bf R}/2|} 
\pm \frac{e^{-\kappa_{\pm}(R)|{\bf r}+{\bf R}/2|}}{|{\bf r}+{\bf R}/2|} \right),
\end{equation}
where ${\cal N}_\pm$ are normalization coefficients which depend on $R$. The corresponding binding energies are  
\begin{equation}\label{BOBindingEnergies}
\epsilon_{\pm} (R) = -\kappa_{\pm}^2(R)/2 m,
\end{equation}
where $m$ is the mass of a light atom, and we put $\hbar=1$. The parameters $\kappa_{\pm}(R)$ follow from the equation
\begin{equation} \label{kappapm}
\kappa_{\pm}(R) \mp \exp \left[-\kappa_{\pm}(R)R \right]/R = 1/a.
\end{equation}
Equation (\ref{kappapm}) is obtained by using the Bethe-Peierls 
boundary conditions for the wavefunctions $\psi^{\pm}_{\bf R}$ of Eq.~(\ref{BOWaveFunctions}) at vanishing light-heavy separations:
\begin{equation} \label{BethePeierls}
\psi^{\pm}_{\bf R} \propto \left(\frac{1}{|{\bf r}\pm {\bf R}/2|}-\frac{1}{a}\right), \quad  |{\bf r}\pm {\bf R}/2| \rightarrow 0.
\end{equation} 

For large $R$ the perturbative expansion of Eq.~(\ref{kappapm}) up to terms of the second order in the small parameter $\exp ( -R/a)$ 
leads to the binding energies
\begin{equation}\label{BOBELargeR}
\epsilon_{\pm} (R) \approx -|\epsilon_0|\mp 2|\epsilon_0|\frac{a}{R}\exp(-R/a)+\frac{U_{\rm ex}(R)}{2},
\end{equation}
where $\epsilon_0=-1/2ma^2$ is the binding energy of a single molecule within the BO approximation with fixed heavy atoms and 
\begin{equation}\label{Exchange}
U_{\rm ex}(R)=4|\epsilon_0|\frac{a}{R}\left(1-\frac{a}{2R}\right)\exp(-2R/a).
\end{equation}
The ungerade ($-$) state energy is always higher than the energy of the gerade ($+$) state. Moreover, for $R<a$ the ungerade state 
is no longer bound. The gerade state is bound for any $R$. For $R\ll a$ its energy is equal to
\begin{equation}\label{GeradeEnergy}
\epsilon_{+} (R)\approx -0.16/mR^2. 
\end{equation}

In the case where the light atoms are noninteracting bosons or distinguishable fermions, it is energetically favorable for them to 
occupy the gerade state and follow it adiabatically when the heavy atoms modify their relative positions. Then the light atoms mediate an 
effective attraction between the heavy atoms. The effective potential in this case is $2\epsilon_{+}(R)+2|\epsilon_0|$, and the remaining 
part of the BO procedure is to solve the Schr\"odinger equation for the heavy atoms moving in this potential. However, such molecules 
are expected to be very short-lived because of the collisional relaxation into deep bound states. Qualitatively, this is similar to the 
situation with homonuclear dimers consisting of bosonic atoms.

From this point on we will focus on the case where light atoms are identical fermions, and light-heavy molecules can be long-lived \cite{petrov_v2_05}.
Then the wavefunction of two light atoms in our BO problem is an antisymmetrized product of the gerade and ungerade wavefunctions
\begin{equation}\label{Antisymm}
\psi_{\bf R}({\bf r_1,r_2})=[\psi_{\bf R}^{+}({\bf r_1})\psi_{\bf R}^{-}({\bf r_2})-\psi_{\bf R}^{+}({\bf r_2})\psi_{\bf R}^{-}({\bf r_1})]/\sqrt{2}.
\end{equation}
The BO adiabatic approach is valid at distances $R>a$, where the effective interaction potential between the molecules is the sum of 
$\epsilon_{+}(R)$ and $\epsilon_{-}(R)$. At sufficiently large inter-heavy separations satisfying the condition $\exp(-R/a)\ll 1$, the effective
potential can be written as 
\begin{equation}\label{BOUeffective}
U_{\rm eff}(R)=\epsilon_{+}(R)+\epsilon_{-}(R)+2|\epsilon_0|\approx U_{\rm ex}(R).
\end{equation} 
The potential $U_{\rm ex}$ originates from the exchange of light fermions and thus can be treated as an exchange interaction. It is purely 
repulsive and, according to Eq.~(\ref{Exchange}), has the asymptotic shape of a Yukawa potential at large $R$. Direct calculations show that $U_{\rm ex}$ 
is a very good approximation to $U_{\rm eff}$ for $R \gtrsim 1.5 a$.
 
We now turn to the second stage of the BO approach and consider the relative motion of two molecules in the center of mass reference frame.
The corresponding Schr\"odinger equation reads
\begin{equation}\label{BOSchr}
( -\nabla^2_{\bf R}/M+U_{\rm eff}(R) -\epsilon)\Psi({\bf R})=0,
\end{equation}
where $\epsilon$ is the collision energy and $M$ is the mass of a heavy atom. Note, that the repulsive effective potential is inversely proportional to 
the light mass $m$, whereas the kinetic energy operator in Eq.~(\ref{BOSchr}) has a prefactor $1/M$. Therefore, for a large mass ratio 
$M/m$, the heavy atoms approach each other at distances smaller than $a$ with an exponentially small tunneling probability $P\propto \exp (-B\sqrt{M/m})$, 
where $B\sim 1$. Our analysis shows that the elastic part of the scattering amplitude can be calculated with a very high accuracy from Eq.~({\ref{BOSchr}) 
for $M/m\gtrsim 20$ and is practically insensitive to the way we choose the boundary condition for the wavefunction at $R=a$. 

When the heavy atoms are fermions one has bosonic light-heavy molecules and the dimer-dimer $s$-wave scattering length $a_{dd}$ is of the order 
of $a\ln\sqrt{M/m}$. The effective range of the potential has the same property. 
Let us demonstrate the calculation of $a_{dd}$ for two bosonic dimers in the limit $M/m\gg 1$. In this case the dominant contribution to the scattering 
comes from distances in the vicinity of $R=a_{dd}\gg a$, where the effective potential can be approximated by Eq.~(\ref{Exchange}) with a constant preexponential factor
\begin{equation}\label{BOExp}
U_{\rm eff}(R)\approx 2(ma a_{dd})^{-1}\exp(-2R/a).    
\end{equation}
Then the zero energy solution of Eq.~(\ref{BOSchr}) that decays towards smaller $R$ reads
\begin{equation}\label{BOApproxSol}
\Psi(R) = \frac{a}{R}K_0\left(\sqrt{\frac{2M}{m}\frac{a}{a_{dd}}}e^{-R/a}\right),    
\end{equation}
where $K_0$ is the decaying Bessel function. Comparing the result of Eq.~(\ref{BOApproxSol}) at large $R$ with the asymptotic behavior $\Psi(R) \propto (1-a_{dd}/R)$ 
we obtain an equation for $a_{dd}$:
\begin{equation}\label{add}
a_{dd}=\frac{a}{2}\ln \left(\frac{e^{2\gamma}}{2} \frac{M}{m}\frac{a}{a_{dd}}\right).
\end{equation}
This gives $a_{dd}\approx a\ln \sqrt{M/m}$, and the scattering cross section is
\begin{equation}    \label{sigmadd}
\sigma_{dd}=8\pi a_{dd}^2.
\end{equation}
From Eq.~(\ref{BOApproxSol}) we see that the interval of distances near $R=a_{dd}$, where the wavefunction changes, is of the order of $a$. This justifies 
the use of Eq.~(\ref{BOExp}). In fact, the corrections to Eq.~(\ref{add}) can be obtained by treating the 
difference between Eqs.~(\ref{Exchange}) and (\ref{BOExp}) perturbatively. In this way the first order correction to the dimer-dimer scattering length is 
$-(3/4)a^2/a_{dd}$, where $a_{dd}$ is determined from Eq.~(\ref{add}).

Qualitatively, $U_{\rm eff}(R)$ can be viewed as a hardcore potential with the radius $a_{dd}$, where the edge is smeared out on a lengthscale 
$\sim a\ll a_{dd}$. Therefore, the ultracold limit for dimer-dimer collisions, required for the validity of Eq.~(\ref{sigmadd}), is realized for relative
momenta of the dimers, $k$, satisfying the inequality
\begin{equation}   \label{ul}
ka_{dd}\ll 1.
\end{equation}
It can be useful (see Ref.~\cite{PAPSS}) to approximate the potential $U_{\rm eff}$ by a pure 
hardcore with the radius $a_{dd}$. This approximation works under the condition $ka\ll 1$, which is less strict than Eq.~(\ref{ul}).

In the case where the heavy atoms are bosons, the diatomic molecules formed by these atoms with the light fermions are composite fermions and they can scatter from each other
only with odd orbital angular momenta. Then, at low momenta $k$ satisfying the condition of the ultracold limit (\ref{ul}), the leading channel is the $p$-wave 
scattering. At interdimer separations exceeding the radius $a_{dd}$ of the interaction potential $U_{\rm eff}(R)$, but still smaller than their 
de Broglie wavelength $1/k$, the radial wavefunction of the relative motion of the dimers takes the form $\Psi(R)\propto k^2(R-3\beta_{dd}/R^2)$, where the quantity
$\beta_{dd}$ is the $p$-wave scattering volume. The scattering amplitude then reads
\begin{equation}    \label{ap}
f_{dd}^{(p)}=k^2\beta_{dd}.
\end{equation}
For the hardcorelike potential $U_{\rm eff}(R)$ one finds that the scattering volume is given by:
\begin{equation}    \label{betadd}
\beta_{dd}\approx (1/3)a_{dd}^3,
\end{equation} 
with $a_{dd}$ following from Eq.~(\ref{add}). Accordingly, the scattering cross section is
\begin{equation}   \label{sigmapdd}
\sigma_{dd}^{(p)}\approx (8\pi/3)k^4a_{dd}^6.
\end{equation}

Note that equations (\ref{add}) and (\ref{betadd}) are obtained in the extreme limit $M\gg m$ and assuming the absence of inelastic processes.
The accuracy of these results for not extremely large ratios $M/m$ and their possible modifications due to the presence of inelastic scattering channels
will be discussed in Sec.~\ref{V}.

Let us now mention that in our numerical calculations presented in Sec.~\ref{V} we do not find resonances in the dimer-dimer scattering amplitude, which
could appear in the presence of a weakly bound state of two dimers. Here we give a qualitative explanation of the absence of these bound states. Suppose 
there is such a state with energy $\epsilon\rightarrow 0$. Then, at distances $R>a$ the wavefunction of the heavy atoms should exponentially decay on the 
distance scale $\sim a\sqrt{m/M}\ll a$, since $U_{\rm eff}$ represents a barrier with the height $\sim 1/ma^2$. This means that the heavy atoms in such 
a bound state should be localized mostly at distances smaller than $a$. The gerade light atom is also localized at these distances as seen from the shape 
of the function $\psi^{+}$. The motion of the ungerade light atom relative to the localized trimer can be viewed as scattering with odd values of 
the angular momentum, and due to the centrifugal barrier the bound states of this atom with the trimer should be localized at distances $\sim a$
from the heavy atoms. In this case one would expect the BO approximation to work, since the ungerade light atom is moving much faster than the heavy 
atoms. However, this leads to a contradiction, because in the BO approach discussed above the ungerade state at interheavy separations $R<a$ is unbound.
We thus conclude that weakly bound states of two dimers are absent.

Although there are no resonances in the dimer-dimer collisions, there are branch-cut singularities in the scattering amplitude. They are related to the presence
of inelastic processes and are discussed in the next section.

\section{Inelastic processes. Qualitative analysis}\label{III}
 
We distinguish two types of inelastic processes in dimer-dimer collisions. The first one is the relaxation of one of the colliding dimers into a deep bound state, 
the other dimer being dissociated. The second process is the formation of bound trimers consisting of two heavy and one light atom, the other light atom carrying away 
the released binding energy. We will start with the formation of trimer states, which in most cases can be called Efimov trimers. The existence of such trimer states 
can be seen from the BO picture for two heavy atoms and one light atom in the gerade state. Within the BO approach the three-body problem reduces to the calculation 
of the relative motion of the heavy atoms in the effective potential created by the light atom. For the light atom in the gerade state, this potential is $\epsilon_+(R)$ 
found in the previous section. The Schr\"odinger equation for the wavefunction of the relative motion of the heavy atoms $\chi_{\nu}({\bf R})$ reads
\begin{equation}\label{BO3body}
\hat H\chi_\nu ({\bf R})=\left[ -\nabla^2_{\bf R}/M+\epsilon_{+}(R)\right]\chi_\nu({\bf R}) =\epsilon_\nu\chi_\nu({\bf R}).
\end{equation}
The trimer states are nothing else than the bound states of heavy atoms in the effective potential $\epsilon_{+}(R)$. Accordingly, they correspond to the discrete part 
of the spectrum $\epsilon_{\nu}$, where the symbol $\nu$ denotes a set containing angular ($l$) and radial ($n$) quantum numbers. For $R\ll a$ the potential $\epsilon_{+}(R)$ 
is proportional to $-1/R^2$ [see Eq.~(\ref{GeradeEnergy})] and, if this effective attraction overcomes the centrifugal barrier, we arrive at the well known phenomenon of the 
fall of a particle to the center in an attractive $1/R^2$ potential. Then, for a given orbital angular momentum $l$, the radial part of $\chi_\nu$ can be written as
\begin{equation} \label{TrimerSmallR}
\chi_\nu(R) \propto R^{-1/2}\sin(s_l \ln{R/r_0}), \quad R \ll a,
\end{equation} 
where 
\begin{equation}\label{sl}
s_l=\sqrt{0.16M/m-(l+1/2)^2}.
\end{equation} 
The three-body parameter $r_0$ fixes the phase of the wavefunction at small distances and, in principle, depends on $l$. The wavefunction (\ref{TrimerSmallR}) has infinitely many 
nodes, which means that in the zero-range approximation there are infinitely many trimer states. This is one of the properties of three-body systems with resonant interactions 
discovered by Efimov \cite{efimov70}. We see that the fall to the center is possible in many angular momentum channels, provided the mass ratio is sufficiently large. However, 
for practical purposes and for simplicity, it is sufficient to consider the case where the Efimov effect occurs only for the angular momentum channel with the lowest possible 
$l$ for a given symmetry. This implies that when the heavy atoms are fermions and one has odd $l$, in order to confine ourselves to $l=1$ we should have the mass ratio in the 
range $14\lesssim M/m \lesssim 76$. For bosonic heavy atoms where $l$ is even, we set $l=0$ and consider $M/m\lesssim 39$ to avoid the Efimov effect for $l\geq 2$. 
In both cases we need a single three-body parameter $r_0$.

The formation of Efimov trimers in ultracold dimer-dimer collisions is energetically allowed only if $\epsilon_\nu<-2|\epsilon_0|$. 
This means that the trimers that we are interested in are relatively well bound and their size is smaller than $a$. Therefore, the process of the trimer formation is exponentially 
reduced for large mass ratios as the heavy atoms have to tunnel under the repulsive barrier $U_{\rm eff}$(R). Moreover, this process requires all of the four atoms to approach 
each other at distances smaller than $a$, and its rate decreases with the trimer size because it is more difficult for two identical light fermions to be in a smaller 
volume. In order to carry out the quantitative analysis of the trimer formation one should properly take into account the motion of the ungerade light fermion at $R<a$. A method 
which allows one to do this will be presented in Sec.~\ref{IV}.

From Eq.~(\ref{TrimerSmallR}) one sees that the behavior of the three-body system does not change if $r_0$ is multiplied by
\begin{equation}\label{lambda}
\lambda_l=\exp(\pi /s_l).
\end{equation} 
On the other hand, the dimensional analysis shows that the quantity $\epsilon_\nu/\epsilon_0$ depends only on the ratio $a/r_0$. This means that except for a straightforward
scaling with $a$, properties of the three-body system do not change when $a$ is multiplied or divided by $\lambda_l$. This discrete scaling symmetry of a three-body system, 
which shows itself in the log-periodic dependence of three-body observables, has yet to be observed experimentally. In the case of three identical bosons, where the BO approach
does not work and one solves the three-body problem exactly \cite{efimov70}, the observation of the consequences of the discrete scaling requires to change $a$ by a factor of 
$\lambda\approx 22.7$, which is technically very difficult in ongoing experiments with cold atoms. In this respect three-body systems with a very large mass difference 
can be more favorable because of smaller values of $\lambda$. For example, in order to see one period of the log-periodic dependence in a Cs-Cs-Li three-body system $a$ has 
to be changed only by a factor of $\lambda \approx 5$.

At this point it is worth emphasizing that three-body effects can be observed in a gas of light-heavy dimers, where the interdimer repulsion originating from the exchange 
of the light fermions strongly reduces the decay rate associated with relaxation of the dimers into deep bound states. 
The trimer formation in dimer-dimer collisions is very sensitive to the positions and sizes of Efimov states, and the measurement of the formation rate can be used for 
demonstrating the discrete scaling symmetry of a three-body system. Indeed, this rate should have the log-periodic dependence on $a$ and is detectable by measuring the
lifetime of the gas of dimers. 

Besides the Efimov trimers, one can have ``universal'' trimer states of one light and two heavy atoms, well described in the zero-range approximation without introducing the 
three-body parameter \cite{Kartavtsev1,Kartavtsev2}. In particular, they exist at the orbital angular momentum $l=1$ and mass ratios below the critical value, where the Efimov 
effect is absent and short-range physics drops out of consideration. One of such states emerges at $M/m \approx 8$ and crosses the trimer formation threshold 
($\epsilon_{tr}=-2|\epsilon_0|$) at $M/m\approx 12.7$. The existence of this state is already seen in the BO picture. It appears 
as a bound state of fermionic heavy atoms in the potential $\epsilon_+(R)$ for $l=1$, and in Sec.~\ref{V} we present estimates for the formation rate in dimer-dimer collisions.
The other state exists at $M/m$ even closer to the critical mass ratio and never becomes sufficiently deeply bound to be formed in cold dimer-dimer collisions. The universal
trimer states also exist for $l>1$ and $M/m>13.6$ \cite{Kartavtsev2}. However, the trimer formation in dimer-dimer collisions at such mass ratios is dominated by the contribution
of Efimov trimers with smaller $l$. Therefore, below we focus on the formation of Efimov trimers. 

Let us now discuss the relaxation of the dimers into deep bound states in dimer-dimer collisions. The typical size of a deep bound state is of the order of the characteristic
radius of
the corresponding interatomic potential. We first consider the relaxation channel that requires one light and two heavy atoms to approach each other to distances $\sim \tilde
R_e\ll a$. 
Unlike the trimer formation, this decay mechanism is a pure three-body process. The other light atom is just a spectator. A qualitative scenario of this process is the following. 
With the tunneling probability which is exponentially suppressed for large $M/m$, two dimers approach each other at distances $R\sim a$. Then the heavy atoms are accelerated 
towards each other in the potential $\epsilon_{+}(R)$, and the light atom in the gerade state is always closely bound to the heavy ones as is seen from the shape of the function 
$\psi^{+}$. The most convenient way to take into account the relaxation process that occurs when the heavy atoms (and the gerade light fermion) are at interatomic separations 
$\sim \tilde R_e$, is to consider the three-body parameter $r_0$ as a complex quantity and introduce the so-called elasticity parameter 
$\eta_*=-s_l \mathrm{Arg} (r_0)$ \cite{braaten07}. As follows from the asymptotic expression for the wavefunction (\ref{TrimerSmallR}), 
a negative argument of $r_0$ ensures that the incoming flux of heavy atoms is not smaller than the outgoing one:
\begin{equation}\label{Imr0}
\Phi_{\rm out}/\Phi_{\rm in}=\exp[4 s_l \mathrm{Arg} (r_0)] = \exp[-4\eta_*]\leq 1.
\end{equation}
This mimics the loss of atoms at small distances due to the relaxation into deep bound states. In the analysis of Efimov states, the imaginary part of $r_0$ leads to 
the appearance of an imaginary part of $\epsilon_\nu$. This means that any Efimov state has a finite lifetime $\tau$ due to the relaxation. 
For small $|\mathrm{Arg} (r_0)|$ and for trimer states that are localized at distances smaller than $a$, we get $\tau^{-1}/|\epsilon_{\nu}|=4|\mathrm{Arg} (r_0)|=4\eta_*/s_l$. 
Strictly speaking this fact indicates that it is not possible to separate the relaxation process from the trimer formation  
because the trimers that are formed in dimer-dimer collisions will eventually decay due to the relaxation. Nevertheless, as we will show in Sec.~\ref{V}, 
both the modulus and argument of the three-body parameter can be determined by measuring the lifetime of a gas of dimers, leading to a number of quantitative 
predictions concerning the structure of Efimov states in the three-body subsystem of one light and two heavy atoms.

Another relaxation channel is the one in which two light fermions approach a heavy one at distances $\sim R_e\ll a$.
This decay mechanism is quite different from the ones described above. On the one hand, it is not exponentially suppressed for large mass ratios as the heavy atoms do not 
have to approach each other. However, there is a suppression of the decay because of the Fermi statistics for the light atoms, which strongly reduces the probability 
of having them in a small volume.
We establish the dependence of the relaxation rate constant on $a$ and estimate its dependence on $R_e$ and the atom masses without performing a detailed analysis of 
the short-range physics. For a fixed and large inter-heavy separation $R$, the relaxation rate $\Gamma (R)$ is proportional to the probability $W$ of finding two light atoms within
a sphere of radius $\sim R_e$ 
around the heavy atom, multiplied by the frequency of relaxation events. This frequency is of the order of $1/mR_e^2$ since there is no other energy scale at distances $\sim R_e$. 
The probability $W$ can be obtained considering the limit ${\bf r_1,r_2}\rightarrow -{\bf R}/2$ in Eq.~(\ref{Antisymm}). To the leading order in $\exp(-R/a)$ we get
\begin{equation}\label{LimAntisymm}
\psi_{\bf R}({\bf r_1,r_2})\approx \frac{e^{-R/a}}{2^{3/2}\pi a^2 R} \left( 1 +
\frac{a}{R} \right) \left[ \frac{\bf \tilde r_2 \cdot R}{\tilde r_1 R} - \frac{\bf \tilde r_1 \cdot R}{\tilde r_2 R}\right],
\end{equation} 
and $W\sim |\psi_{\bf R}({\bf r_1,r_2})|^2 R_e^6$ with $\tilde r_{1,2}=|{\bf r_{1,2}}+{\bf R}/2 |\sim R_e$ in the expression for the wavefunction $\psi_{\bf R}$. This gives an estimate for the relaxation 
rate $\Gamma(R)$ as a function of $R$:
\begin{equation}\label{Gamma}
\Gamma(R)\sim (1/mR_e^2)(R_e^6/a^4R^2)\exp(-2R/a).
\end{equation}
Finally, $\Gamma (R)$ should be averaged over the motion of the heavy atoms, which leads to the relaxation rate constant for the heavy-light-light process in dimer-dimer
collisions:
\begin{equation}\label{Alphahll}
\alpha_{hll} = \int \Gamma(R)|\Psi(R)|^2 {\rm d}^3 R \sim \frac{R_e^4}{Ma^3}\frac{1}{\ln\sqrt{M/m}}.
\end{equation}
The main contribution to the integral in Eq.~(\ref{Alphahll}) comes from distances $R \gg a$ where we use Eq.~(\ref{BOApproxSol}) for the wavefunction $\Psi(R)$. 
From Eq.~(\ref{Alphahll}) we see that $\alpha_{hll}\propto a^{-3}$, i.e. this relaxation mechanism is strongly suppressed for large dimers. 

In the case of fermionic molecules, i.e. $p$-wave collisions, the relaxation rate constant (\ref{Alphahll}) should be multiplied by an additional factor 
$(ka)^2\ln^2\sqrt{M/m}$, where $k$ is the scattering momentum.

Summarizing the results of sections \ref{II} and \ref{III} we find that the simple BO approach works well for calculating the elastic part of the dimer-dimer scattering 
amplitude as the main contribution to this quantity comes from large distances between the heavy atoms, where the adiabatic approximation for the motion of light atoms is valid.
This is also the 
case for the relaxation into deep bound states which occurs in the three-body subsystem consisting of one heavy and two light atoms. The other inelastic processes (trimer formation
and 
the relaxation occurring in the system of one light and two heavy atoms) require two heavy atoms to approach each other to distances $R<a$, where the adiabatic approximation 
breaks down. Quantitative studies of these inelastic processes will provide us with the estimate for the lifetime of the gas of dimers. These studies are also needed to 
establish the relation between three- and four-body observables, which can be useful for the exploration of three-body effects and Efimov physics. 
In the next section we present a modified BO approach to tackle these issues.

\section{Hybrid Born-Oppenheimer method\label{IV}}

As we argued in Sec.~\ref{II}, the BO method works well for the gerade fermion. Its state is characterized by the wavefunction $\psi_{\bf R}^{+}({\bf r})$ and 
energy $\epsilon_{+}(R)$, both adiabatically adjusting themselves to the motion of the heavy atoms. In the usual BO approximation one integrates out the light atom by 
introducing a potential energy surface (PES) for the heavy atoms, which in this particular case is given by $\epsilon_{+}(R)$. One can improve the PES by 
incorporating the so-called adiabatic Born-Oppenheimer correction. The idea of this procedure is to retain $\psi_{\bf R}^{+}({\bf r})$ as the wavefunction of the 
light atom, but to calculate the PES as the expectation value of the full Hamiltonian of the three-body system:
\begin{equation}\label{Ham3body}
\hat{H}_3=-\nabla_{\bf R}^2/M-\nabla_{\bf r}^2/2\mu_3+{\rm Int.terms}.
\end{equation}  
Here $\mu_3=2mM/(2M+m)$, and ${\bf r}$ is the coordinate of the light atom relative to the center of mass of the heavy atoms. The interaction terms describe the 
short range light-heavy interactions equivalent to the boundary conditions (\ref{BethePeierls}). The PES calculated by using $\psi_{\bf R}^{+}({\bf r})$ of
Eq.~(\ref{BOWaveFunctions}) reads: 
\begin{equation}\label{PES}
\langle \psi^{+} |\hat{H}_3|\psi^{+}\rangle = -\kappa_{+}^2(R)/2\mu,
\end{equation} 
where $\mu=mM/(M+m)$. It is easy to see that in our case the adiabatic correction results in the substitution $m\rightarrow \mu$ in all formulas of the 
previous sections. In the following we will keep this substitution in mind and, in particular, use the same notations $\epsilon_{\pm}(R)$ and $\epsilon_0$ for the ``renormalized''
quantities. 

Strictly speaking, the adiabatic correction does not improve the accuracy of the BO procedure as we are still missing non-adiabatic terms which are of the same order of magnitude. 
However, including this correction has the advantage that the renormalized quantities $\epsilon_{\pm}(R)$ asymptotically tend to the exact binding energy of a dimer, 
$\epsilon_0=-1/2\mu a^2$, which is more convenient for further analysis. 

Once the light atom in the gerade state is integrated out, the original four-body problem reduces to a three-body problem, which is described by the three-body Schr\"odinger
equation
\begin{equation}\label{Reduced3body}
[\hat{H}-\nabla_{\bf r}^2/2\mu_3-E]\Psi({\bf R},{\bf r})=0.
\end{equation}
Here $\hat{H}$ is given by Eq.~(\ref{BO3body}), $E=-2|\epsilon_0|+\epsilon$ is the total energy of the four-body system in the center of mass reference frame, 
and $\epsilon$ is the dimer-dimer collision energy. The interaction of the light atom with the heavy ones is included in the form of the boundary condition (\ref{BethePeierls}) 
for $\Psi$, and the ungerade symmetry for this atom is taken into account by the condition 
\begin{equation}\label{CondUngerade}
\Psi({\bf R},{\bf r})=-\Psi({\bf R},-{\bf r}).
\end{equation}

Let us now discuss the symmetry aspects of the reduced three-body wavefunction $\Psi({\bf R},{\bf r})$. In the case where the heavy atoms are identical fermions, 
we have $\Psi({\bf R},{\bf r})=-\Psi(-{\bf R},{\bf r})$. Combined with Eq.~(\ref{CondUngerade}), this leads to the condition 
$\Psi({\bf R},{\bf r})=\Psi(-{\bf R},-{\bf r})$. Therefore, $\Psi({\bf R},{\bf r})$ describes atom-dimer scattering with even angular momenta, and for 
ultracold collisions we are facing the $s$-wave atom-dimer scattering problem. Performing a similar analysis for the case where the heavy atoms are identical 
bosons we naturally arrive at the $p$-wave atom-dimer scattering problem. Both cases are consistent with our understanding of the dimer-dimer scattering once we 
recall that there exists a ``hidden'' gerade fermion. 

In order to solve Eq.~(\ref{Reduced3body}) we follow the method of Ref. \cite{petrov03}. Namely, we introduce an auxiliary function $f({\bf R})$ and write down the 
wavefunction $\Psi({\bf R},{\bf r})$ in the form:
\begin{equation} \label{fourbodysolution}
\Psi({\bf R,r})=\sum_{\nu}\int_{\bf R'} \chi_{\nu}({\bf R}) \chi_{\nu}^{\ast}({\bf R'})K_{\kappa_\nu}(2 {\bf r},{\bf R'})f({\bf R'}),
\end{equation}
where
\begin{equation}\label{Knu}
K_{\kappa_\nu}(2 {\bf r},{\bf R'})=\frac{e^{-\kappa_\nu |{\bf r}-{\bf R'}/2|}}{4\pi |{\bf r}-{\bf R'}/2|}-\frac{e^{-\kappa_\nu |{\bf r}+{\bf R'}/2|}}{4\pi |{\bf r}+{\bf R'}/2|}
\end{equation}
and
\begin{equation}\label{kappa}
\kappa_\nu=\left\{ \begin{array}{ll} \sqrt{2\mu_3(\epsilon_\nu-E)},& \epsilon_\nu>E,\\
-i\sqrt{2\mu_3(E-\epsilon_\nu)},& \epsilon_\nu<E. \end{array} \right.
\end{equation}

For $\epsilon_\nu<E$ trimer formation in the state $\nu$ is possible. In this case $\kappa_\nu$ is imaginary and the function (\ref{Knu}) describes an outgoing wave of 
the light atom moving away from the trimer. The choice of the sign in Eq.~(\ref{kappa}) ensures that there is no incoming flux in the atom-trimer channel.  

One can check directly that the wavefunction (\ref{fourbodysolution}) satisfies Eqs.~(\ref{Reduced3body}) and (\ref{CondUngerade}) for any $f({\bf R})$. It is also easy 
to see that for the case of fermionic heavy atoms, where $\Psi({\bf R,r})=-\Psi({\bf -R,r})$, one has $f({\bf R})=f(-{\bf R})$. Then, only antisymmetric $\chi_\nu$ 
are contributing to the sum in Eq.~(\ref{fourbodysolution}). Otherwise, if we choose antisymmetric $f({\bf R})$, it gives $\Psi({\bf R,r})=-\Psi({\bf -R,r})$ and describes bosonic 
heavy atoms with symmetric $\chi_\nu$. 

The function $f({\bf R})$ can be obtained by using the Bethe-Peierls boundary condition (\ref{BethePeierls}) for the wavefunction (\ref{fourbodysolution}). 
Namely, one has to take the limit ${\bf r}\rightarrow {\bf R}/2$ in Eq.~(\ref{fourbodysolution}) and fix the ratio between the regular term and the coefficient in front of the
singular 
term proportional to $1/|{\bf r}- {\bf R}/2|$. In this way one finds an integral equation for $f$. However, this equation is not convenient for numerical calculations, and
we modify it by using the following procedure. 

We note that taking a finite number of terms in the sum over $\nu$ in Eq.~(\ref{fourbodysolution}) does not give any singularities. 
Therefore, the $|{\bf r}- {\bf R}/2|^{-1}$-peak in $\Psi({\bf R,r})$ is provided by the high-energy part of the sum, and the corresponding coefficient does not depend 
on the shape of the potential $\epsilon_{+}(R)$ in the Hamiltonian $\hat{H}$. Let us introduce the function $\Psi^0({\bf R},{\bf r})$ by the formula
\begin{equation} \label{fourbodysolution0}
\Psi^0({\bf R,r})=\sum_{\nu}\int_{R'} \chi_{\nu}^0({\bf R}) \chi_{\nu}^{0\ast}({\bf R'})K_{\kappa_\nu^0}(2 {\bf r},{\bf R'})f({\bf R'}),
\end{equation}
where the superscript $0$ in the right hand side corresponds to the eigenfunctions and eigenenergies of Eq.~(\ref{BO3body}) with $\epsilon_{+}(R)\equiv \epsilon_0$. 
The difference $\Psi({\bf R},{\bf r})-\Psi^0({\bf R},{\bf r})$ is not singular for ${\bf r}\rightarrow {\bf R}/2$ and its value in this limit is given by
\begin{widetext}
\begin{equation}\label{Lprime}
\hat{L}'f({\bf R})=\Psi({\bf R},{\bf R}/2)-\Psi^0({\bf R},{\bf R}/2)=\int_{\bf R'} \sum_{\nu} [\chi_{\nu}({\bf R}) \chi_{\nu}^{\ast}({\bf R'})K_{\kappa_\nu}({\bf
R,R'})-\chi_{\nu}^0({\bf R}) 
\chi_{\nu}^{0\ast}({\bf R'})K_{\kappa_\nu^0}({\bf R,R'})]f({\bf R'}).
\end{equation}

The function $\Psi^0({\bf R},{\bf r})$ can be calculated straightforwardly by turning the sum over $\nu$ in Eq.~(\ref{fourbodysolution0}) into an integral over ${\bf k}$:
\begin{equation}
\sum_{\nu} \chi_{\nu}^0({\bf R}) \chi_{\nu}^{0\ast}({\bf R'}) \rightarrow (2\pi)^{-3} \int {\rm d}^3k e^{i{\bf k\cdot (R-R')}},
\end{equation}
and substituting $\epsilon_{\nu}^0 \rightarrow k^2/M+\epsilon_0$. In the limit of ${\bf r} \rightarrow {\bf R}/2$ we get
\begin{equation} \label{threebodyeq}
\Psi^0({\bf R},{\bf r}\rightarrow {\bf R}/2)=\sin^2{\theta} ( |{\bf r}-{\bf R}/2|^{-1}-\sqrt{2\mu(\epsilon_0-E)} )f({\bf R})/4\pi-\hat{L} f({\bf R}),
\end{equation}
where $\theta=\arctan{\sqrt{1+2M/m}}$, and the integral operator $\hat{L}$ is given by
\begin{equation}\label{L}
\hat{L} f({\bf R}) = P\int_{\bf R'} \left\{ G(|{\bf R-R'}|)\left[ f({\bf R})-f({\bf R'}) \right]\pm G(\sqrt{R^2+R'^2-2{\bf R\cdot R'}\cos{2\theta}})f({\bf R'}) \right\},
\end{equation}
\end{widetext}
with the symbol $P$ indicating the Principal Value of the integral (see \cite{petrov05}). The plus (minus) sign in Eq.~(\ref{L}) corresponds to the case of fermionic (bosonic)
heavy atoms. 
The function $G$ is defined as
\begin{equation}
G(X)=\frac{\sin 2\theta M (\epsilon_0-E) K_2(\sqrt{M (\epsilon_0-E)}X/\sin\theta)}{8\pi^3 X^2},
\end{equation}
where $K_2(z)$ is the exponentially decaying Bessel (Macdonald) function.

Finally, making a summation of Eq.~(\ref{Lprime}) with Eq.~(\ref{threebodyeq}) and comparing the result with the boundary condition (\ref{BethePeierls}), we find the 
following equation for $f({\bf R})$:
\begin{equation}\label{final}
\left\{\hat{L}-\hat{L}'+\sin^2\theta\frac{\sqrt{2\mu(\epsilon_0-E)} -1/a}{4\pi}\right\}f({\bf R})=0.
\end{equation}
The operators $\hat{L}$ and $\hat{L}'$ conserve angular momentum and, expanding the function $f({\bf R})$ in spherical harmonics, we deal with a set of uncoupled 
1D integral equations for each of the radial functions $f_l(R)$. These equations for $l=0$ and $l=1$ are given in Appendix.

\section{Low-energy dimer-dimer scattering}\label{V}

The hybrid Born-Oppenheimer approximation (HBO) developed in Sec.~\ref{IV} can be used in various ways for obtaining the rates of elastic and inelastic dimer-dimer
collisions. For example, substituting the solution of Eq.~(\ref{final}) into Eq.~(\ref{fourbodysolution}) one can, in principle, restore the total wavefunction $\Psi({\bf R},{\bf
r})$
and find the scattering amplitudes from the asymptotic shape of this wavefunction at $R\rightarrow\infty$. 
However, for the problem of ultracold dimer-dimer scattering this is not needed as there is a more elegant procedure. 

At large distances ($R\gg a$) the reduced wavefunction $\Psi({\bf R},{\bf r})$ takes the form:
\begin{equation}\label{asymptote}
\Psi({\bf R},{\bf r})\approx\Psi({\bf R})\psi_{\bf R}^{-}({\bf r}).
\end{equation}
Comparing the singular parts of Eqs.~(\ref{threebodyeq}) and (\ref{asymptote}) in the limit of ${\bf r} \rightarrow {\bf R}/2$ we obtain
\begin{equation}\label{Psif}
f({\bf R})\propto \Psi({\bf R});\,\,\,\,\,\, R\gg a.
\end{equation}
Thus, at large distances $f({\bf R})$ can serve as the wavefunction for the dimer-dimer motion. In particular, it contains the dimer-dimer scattering phase shift.

In the HBO approach the dimer-dimer scattering amplitude $a_{dd}$ is determined from the large distance behavior of the solution of Eq.~(\ref{final}) 
for $E=2\epsilon_0$. In the case of fermionic heavy atoms, the leading channel at ultralow energies is the $s$-wave scattering, 
and the solution of Eq.~(\ref{final}) should be matched with   

\begin{equation}\label{Psifs}
f_0(R)\propto (1/R-1/a_{dd})
\end{equation}
at $R\gg a\ln \sqrt{M/m}$. In the case of the $p$-wave dimer-dimer scattering (bosonic heavy atoms) one should match the solution of Eq.~(\ref{final}) 
at these distances with
\begin{equation}\label{Psifp}
f_1(R)\propto (R-3\beta_{dd}/R^2). 
\end{equation}
The numerical procedure of HBO is described below. The dotted curve in Fig.~\ref{ascat} represents $a_{dd}/a$ from Eq.~(\ref{add}) and $(3\beta_{dd})^{1/3}/a$ extracted from 
Eq.~(\ref{betadd}). The HBO results for these quantities are shown by the solid and dashed curves, respectively. One sees a relatively good agreement even for moderate values
of $M/m$. The HBO result for $a_{dd}$ agrees with our previous calculations which use the exact four-body equation for $M/m<13.6$ (see \cite{petrov_v2_05} and Sec.~\ref{VI}). 
We also get a good agreement with the Monte Carlo results for $M/m<20$ \cite{Blume}.
 
\begin{figure}[h]
\includegraphics[width=\hsize,clip]{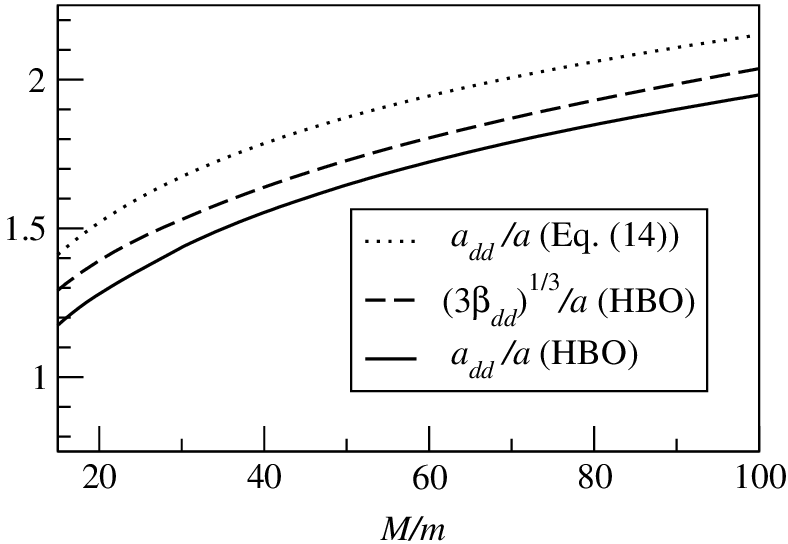}
\caption{
The dimer-dimer $s$-wave scattering length $a_{dd}/a$ (solid) and the quantity $(3\beta_{dd})^{1/3}/a$ for $p$-wave scattering (dashed) calculated by using HBO. 
The dotted line is the result of Eq.~(\ref{add}). 
\label{ascat}}
\end{figure}

In this formalism it is straightforward to account for inelastic processes of trimer formation and the relaxation of dimers into deep bound states. In the latter case we mean
the relaxation process that requires one light and two heavy atoms to closely approach each other. The heavy-light-light relaxation channel is well described by the simple BO
approach 
(see Sec.~\ref{III}) and we do not consider it here.

Let us first assume that the rate of relaxation into deep molecular states is negligible and omit this process. Then the three-body parameter is real, and the trimer formation rate
is 
determined by the imaginary part of the $s$-wave scattering length or the $p$-wave scattering volume. The rate constant is given by \cite{LLQ} 
\begin{equation}\label{Inelastic}
\alpha = -\frac{16\pi}{M}\times\left\{ \begin{array}{ll} \mathrm{Im}(a_{dd}),& \textrm{heavy fermions}, \\
3k^2\mathrm{Im}(\beta_{dd}),& \textrm{heavy bosons}. \end{array} \right.
\end{equation}
Alternatively, if one needs to know the rate of trimer formation in the state $\nu$, one can substitute the solution of Eq.~(\ref{final}) into Eq.~(\ref{fourbodysolution}) 
and calculate the flux of light atoms at ${\bf r}\rightarrow \infty$. Summing over $\nu$ gives the same result as Eq.~(\ref{Inelastic}). We find that the contribution of 
the highest ``dangerous'' trimer state is by far dominant and $\alpha$ is very sensitive to its position.

We now include the relaxation of the dimers into deep bound states. 
As we have mentioned in Sec.~\ref{III}, the light-heavy-heavy relaxation process can be taken into account by adding an imaginary part to the three-body parameter. 
The total inelastic decay rate is then still given by Eq.~(\ref{Inelastic}). However, strictly speaking we can no longer distinguish between the trimer formation in a 
particular state and the relaxation since the trimers ultimately decay due to the relaxation process. In this sense the only decay channel is the relaxation. However, for a 
sufficiently long lifetime of a trimer, i.e. if the trimer states are narrow resonances, we can still see a pronounced dependence of the total inelastic decay rate on the 
position of the highest ``dangerous'' trimer state (see below). 

The calculation of the intrinsic lifetime of a trimer requires a detailed knowledge of short-range physics and is beyond the scope of this paper.
Estimates of the imaginary part of the trimer energy, $\tau^{-1}$, from the experimental data on Cs$_3$ trimers \cite{kraemer06} show that it is approximately by a factor of 4
smaller than the real part $\epsilon_{\nu}$ (in this case $\eta_*\approx 0.06$). From a general point of view, we do not expect that the trimers with a binding energy
$\epsilon_{\nu}<-2|\epsilon_0|$ are very long-lived. However, one can have relatively narrow resonances, and we perform calculations for various values of the elasticity parameter
$\eta_*$.  

In the case of a complex three-body parameter the Hamiltonian $\hat{H}$ (\ref{BO3body}) combined with the boundary condition (\ref{TrimerSmallR}), no longer represents a Hermitian 
operator. Instead, this operator is equivalent to the Schr\"odinger operator with a complex absorbing potential, and in a real basis set its representation is a complex 
symmetric matrix. Methods of calculating Green functions and other properties of such operators are known in quantum chemistry (see \cite{Santra} and refs. therein). 
The formalism is based on using the canonical symmetric bilinear form instead of the Hermitian inner product. In our case, this results in the formal removal of the 
sign of the complex conjugation for all $\chi_\nu^\ast$ in Sec.~\ref{IV}. This does not lead to any contradiction with the ``Hermitian'' case as the eigenvectors of 
Hermitian operators can be made real.  

We have performed a numerical analysis of the elastic and inelastic rates for bosonic and fermionic heteronuclear dimers for different values of $M/m$. 
A brief sketch of the numerical procedure is the following. We diagonalize the Hamiltonian $\hat{H}$ of Eq.~(\ref{BO3body}) on a finite but large grid in coordinate space and find
the sets 
$\{\epsilon_\nu,\chi_\nu({\bf R})\}$ and $\{\epsilon_\nu^0,\chi_\nu^0({\bf R})\}$. The latter is obtained from Eq.~(\ref{BO3body}) with $\epsilon_{+}(R)\equiv \epsilon_0$. 
The three-body boundary condition (\ref{TrimerSmallR}) for $\chi_\nu$ is implemented by introducing an additional (complex) potential at very short distances. 
Then, we calculate the kernels of the integral operators $\hat{L}$ and $\hat{L}'$ and solve Eq.~(\ref{final}). In order to characterize the elastic and inelastic dimer-dimer 
scattering properties we use Eqs.~(\ref{Psifs}), (\ref{Psifp}), and (\ref{Inelastic}).

In Fig.~\ref{YbLi} we present the results for the inelastic collisional rate in the case of bosonic molecules with the mass ratio $M/m=28.5$ having in mind 
${}^{171}$Yb-${}^6$Li dimers. The solid line corresponds to the case of a real three-body parameter. It is convenient to introduce a related quantity, 
$a_0$, defined as the value of $a$ at which the energy, $\epsilon_\nu$, of a trimer state exactly equals $E=2\epsilon_0$. This new ``dangerous'' trimer state 
becomes more deeply bound for $a>a_0$ and the rate constant steeply grows being proportional to the density of states in the outgoing atom-trimer channel. The corresponding 
orbital angular momentum is equal to 1 and the threshold law reads (see also inset in Fig.~\ref{YbLi})
\begin{equation}\label{Threshold}
\alpha\propto {\rm const}+(E-\epsilon_\nu)^{3/2}\propto {\rm const}+(a-a_0)^{3/2}.
\end{equation}
The constant term in Eq.~(\ref{Threshold}) describes the contribution of more deeply bound states, which is typically very small. In fact, as trimer states 
become more compact, both light atoms should approach the heavy atoms and each other at smaller distances where the trimer formation takes place. Since they are identical fermions, 
there is a strong suppression of the trimer formation to these deeply bound states. 

\begin{figure}[h]
\includegraphics[width=\hsize,clip]{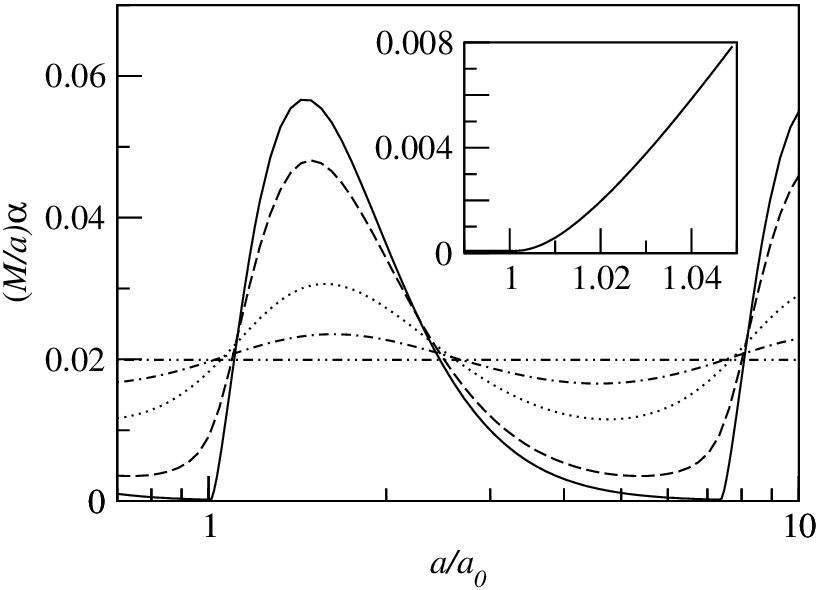}
\caption{
The inelastic rate constant for bosonic dimers with $M/m=28.5$ as a function of the atom-atom scattering length $a$. The solid line corresponds to the case of a real three-body 
parameter. The results plotted in dashed, dotted, dashed-dotted, and dash-dot-dot lines are obtained by taking into account the light-heavy-heavy relaxation processes. The values 
of the elasticity parameter $\eta_*=$0.1, 0.5, 1, and $\infty$, respectively (see text). The quantity $a_0$ is the value of $a$ at which the energy of a trimer state equals 
$E=2\epsilon_0$ and a new inelastic channel opens. The inset shows the region $a\approx a_0$ in greater detail in order to see the threshold behavior (\ref{Threshold}).
\label{YbLi}}
\end{figure}

The plot is periodic in the logarithmic scale, the multiplicative factor being equal to $\lambda_1\approx 7.3$ consistent with Eqs.~(\ref{lambda}) and (\ref{sl}) 
(with $m\rightarrow \mu$). The dashed, dotted, and dashed-dotted curves are obtained for $\eta_*=$0.1, 0.5, and 1, respectively. The corresponding values of the ratio 
$\Phi_{\rm out}/\Phi_{\rm in}$ are 0.67, 0.14, and 0.02. The flat line represents the limiting case of $\eta_*=\infty$ or $\Phi_{\rm out}=0$. This case is universal in 
the sense that physical observables depend only on the masses and the atomic scattering length.

For a very weak light-heavy-heavy relaxation, the dimer-dimer inelastic collision can be viewed as the formation of a trimer (with the rate constant $\alpha$) followed 
by its slow decay due to the relaxation. In this case one can think of detecting the trimers spectroscopically. We note, however, that even for the conditions corresponding 
to the dashed curve in Fig.~\ref{YbLi}, i.e. for $\eta_*$ as small as 0.1, the decay rate of a trimer $\tau^{-1}\approx 0.25|\epsilon_{\nu}|$ is rather fast, which is likely to
make 
its direct detection difficult. For larger $\eta_*$ one cannot separate the formation of trimers from their intrinsic relaxational decay, and $\alpha$ is practically the 
relaxation rate constant. Remarkably, according to our results, it remains quite sensitive to the positions of trimer states (in this case resonances) for values of $\eta_*$ up 
to 0.5 and even larger. This suggests that measuring the lifetime of a gas of dimers as a function of $a$ gives important information on three-body observables. Moreover, 
for small $\eta_*$ one can have a stable molecular gas in sufficiently broad regions of $a$, where ``dangerous'' trimer states are far from the trimer formation threshold.

We have also calculated the inelastic rate $\alpha$ for other mass ratios in the range $20<M/m<76$. Its dependence on the scattering length has the same structure as plotted in 
Fig.~\ref{YbLi}. The maximum of the rate constant is well fitted by the formula $\alpha_{\rm max}= 5.8(a/M)\exp(-0.87\sqrt{M/m})$. The position of the flat line ($\eta_*=\infty$)
is 
also well fitted by $\alpha_{\infty}= 1.6(a/M)\exp(-0.82\sqrt{M/m})$. The multiplicative factor in the log-periodic dependence is given by Eqs.~(\ref{lambda}) and (\ref{sl}) with
$l=1$.

We employed the same method for estimating the formation of the universal trimer state with orbital angular momentum $l=1$ at mass ratios $M/m>12.7$ but below the critical value
for the onset of the Efimov effect. 
The rate constant increases with $M/m$ and reaches $\alpha=0.2(a/M)$ on approach to the critical mass ratio. This corresponds to the imaginary part of
the scattering length ${\rm Im}a_{dd}\approx 4\times 10^{-3}a$, which is by a factor of $300$ smaller than the real part of $a_{dd}$ obtained in four-body calculations
\cite{petrov_v2_05}. Thus, the formation of this state does not change the elastic scattering amplitude $a_{dd}$ of Ref. \cite{petrov_v2_05}.

We can now estimate the rate constants for $^{171}$Yb-$^{6}$Li dimers with the range $R_e < 5$nm. On the basis of Eq.~(\ref{Alphahll}) and the results in 
Fig.~\ref{YbLi} we find that for $a=20$nm the upper bound of the inelastic rate constant is $\alpha_{\rm max}\approx 4\times 10^{-13}$cm$^3/$s and it is larger than 
the rate of relaxation into deep bound states in the system of one heavy and two light atoms, $\alpha_{hll}$. 
For larger $a$ the heavy-light-light relaxation can be neglected and the lifetime of the gas of dimers is determined solely by the rate constant $\alpha$. 
The elastic rate constant for a thermal gas with $a\sim 20$nm and $T\sim 100$nK equals
\begin{equation}\label{Elastic}
\alpha_{el}\approx 8\pi |a_{dd}|^2 \sqrt{2T/M}\sim 4\times 10^{-11}{\rm cm}^3/{\rm s}.
\end{equation}
Here we used the calculated $s$-wave dimer-dimer scattering length $a_{dd}\approx 1.4 a$. We see that $\alpha_{el}$ is much larger than $\alpha$ and this inequality 
becomes even more pronounced for larger $a$ due to the scaling relations  $\alpha_{el}\propto a^2$ and $\alpha_{\rm max}\propto a$.

In Fig.~\ref{RbCsLi} we present the results for the inelastic collisional rate in the case of fermionic dimers with mass ratios 14.5 and 22.2, having in mind  
$^{87}$Rb-$^{6}$Li and $^{133}$Cs-$^{6}$Li dimers respectively. We observe essentially the same structure of the graphs as in the case of bosonic molecules. However, now we have
$p$-wave scattering and 
the rate constant acquires an additional factor $\propto (ak)^2$, where $k$ is the relative momentum of the dimers. The multiplicative factor in the log-periodic dependence is 
now given by Eqs.~(\ref{lambda}) and (\ref{sl}) with $l=0$, since the light-heavy-heavy three-body atomic complex exhibits the Efimov effect in the configuration with zero 
orbital angular momentum. The dimer-dimer scattering volumes are equal to $\beta_{dd}\approx 0.73 a^3$ for $M/m=14.5$ and $\beta_{dd}\approx 0.98 a^3$ for $M/m=22.2$.

\begin{figure}[h]
\includegraphics[width=\hsize,clip]{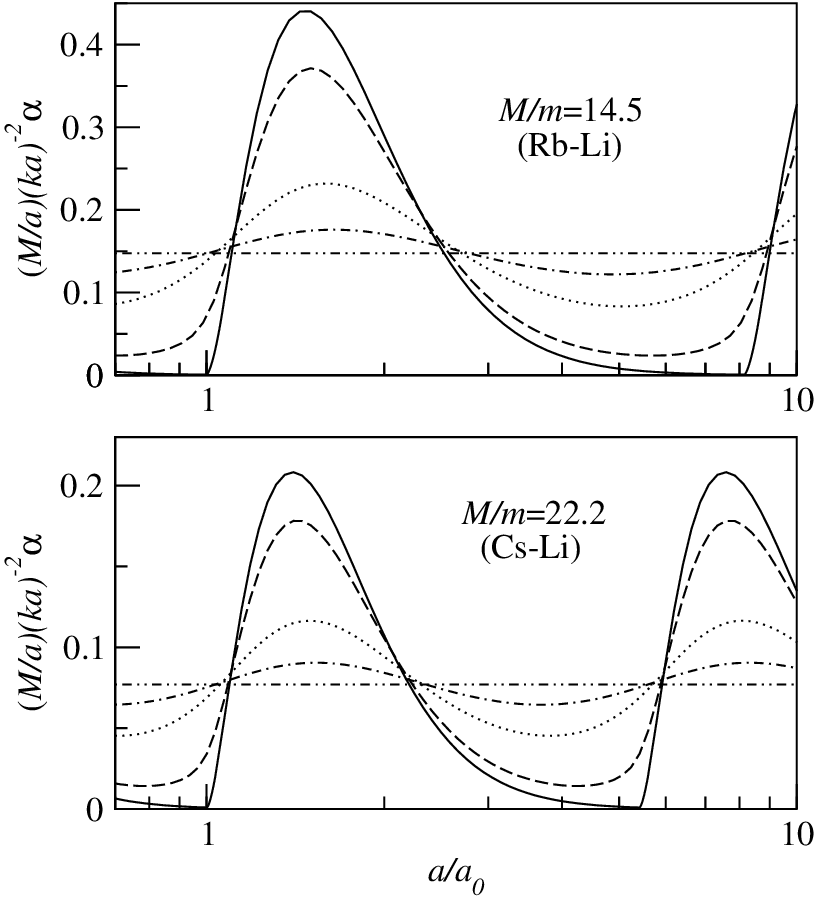}
\caption{
The inelastic rate constant for fermionic dimers with the mass ratios $M/m=14.5$ and $22.2$. Notations are the same as in Fig.~\ref{YbLi}.
\label{RbCsLi}}
\end{figure}

For fermionic molecules with other mass ratios the inelastic rates behave in a similar 
way as plotted in Fig.~\ref{RbCsLi}. Considering $M/m < 39$ in order to avoid the Efimov effect in the $l=2$ channel, the quantities $\alpha_{\rm max}$ and $\alpha_{\infty}$ 
for a given relative momentum $k$ can be well fitted by the formulas $\alpha_{\rm max}= 14.3(a^3k^2/M)\exp(-0.88\sqrt{M/m})$ 
and $\alpha_{\infty}= 2.8(a^3k^2/M)\exp(-0.77\sqrt{M/m})$. The rate constant of elastic collisions is given by
\begin{equation}
\alpha_{el}\approx 48 \pi |\beta_{dd}|^2 k^5/M.
\end{equation}
In the case of Cs-Li dimers we find that for $a\gg R_e\approx 3$nm the upper bound $\alpha_{\rm max}$ greatly exceeds $\alpha_{hll}$. 
Moreover, for a thermal molecular gas with $a=20$nm at $T\sim 100$nK, the quantity $\alpha_{\rm max}$ is comparable with $\alpha_{el}$ and both
are approximately equal to $4\times 10^{-14}$cm$^3/$s. We suggest that in order to optimize the collisional rates 
and lifetime of a gas of dimers and obtain $\alpha_{el}\gg\alpha$ one modifies the scattering length or temperature, since dimer-dimer inelastic and elastic rates 
strongly depend on these parameters. The scaling relations for the rate constants read: $\alpha_{el}\propto a^6k^5$ and $\alpha_{\rm max}\propto a^3k^2$.

It is useful to mention that for fermionic dimers the relaxation rates in the case of atom-dimer collisions are higher than in the case of dimer-dimer collisions. 
The relaxation rate constant for collisions of
Cs-Li dimers with Cs atoms is $\alpha \sim a/m\approx 2\times 10^{-10}$cm$^3/$s and it exceeds by 3 orders of magnitude the relaxation rate for collisions of
these dimers with Li atoms. Although these estimates are valid within an order of magnitude, they indicate that a pure 
gas of dimers should have a longer lifetime. At least, one should avoid mixing Cs-Li dimers with free Cs atoms. 

\section{On the validity of HBO}\label{VI}

Our derivation of the HBO method in Sec.~\ref{IV} is based on two assumptions. First, we rely on the short-range character of the interatomic forces. 
Namely, all our results are valid under the condition (\ref{aRe}). It assumes that the range, $R_e$, of the interatomic van der Waals potential is much smaller 
than the typical de Broglie wavelength of atoms, which is of the order of $a$. In this case we can treat the interatomic potential using the zero-range 
Bethe-Peierls approach \cite{bethe35}. 

The second important assumption is the use of the Born-Oppenheimer approximation for integrating out the light fermion in the gerade state. We rely on the presence of
the small parameter $m/M$. In the beginning of Sec.~\ref{IV} we explained how the simplest (clamped nuclei) approximation can be improved by adding the Born-Oppenheimer 
diagonal correction. The resulting so-called adiabatic approximation provides one with the ``best possible'' potential energy surface for the heavy atoms. 
In principle, one can make further improvements and systematically expand the wavefunctions and energies in powers of $m/M$. However, it is believed that the amount of work 
that one has to do in order to get the next order term in this expansion is comparable to the work required for the exact solution of the problem. 

In such a situation it looks optimal to check the validity of the HBO results by comparing them with the ones obtained in exact four-body calculations \cite{petrov_v2_05}. 
In Fig.~\ref{benchmark} we compare $a_{dd}$ for bosonic molecules in the 3D case, calculated exactly (solid curve) and by using the HBO method (dashed curve). Already for 
$M/m\sim 10$ the results agree within the limits of accuracy. The exact method requires a huge configuration space for $M/m\approx 10$ and its accuracy drops significantly 
at larger mass ratios. We estimate the relative precision of this ``exact'' four-body calculation at the highest mass ratio in Fig.~\ref{benchmark} to be 2 or 3$\%$, whereas
one should have a much higher accuracy in order to resolve the trimer formation for $M/m>12.7$.  
In fact, we believe that the HBO method is more accurate for $M/m\gtrsim 10$. To justify this belief we have performed the same analysis in the 1D case, 
where the exact method can be pushed to higher mass ratios. The HBO results agree very well with the results of the exact methods (see inset in Fig.~\ref{benchmark}). 
In the 1D case for the considered mass ratios there are no trimer states and we neglect any inelastic decay.

\begin{figure}[h]
\includegraphics[width=\hsize,clip]{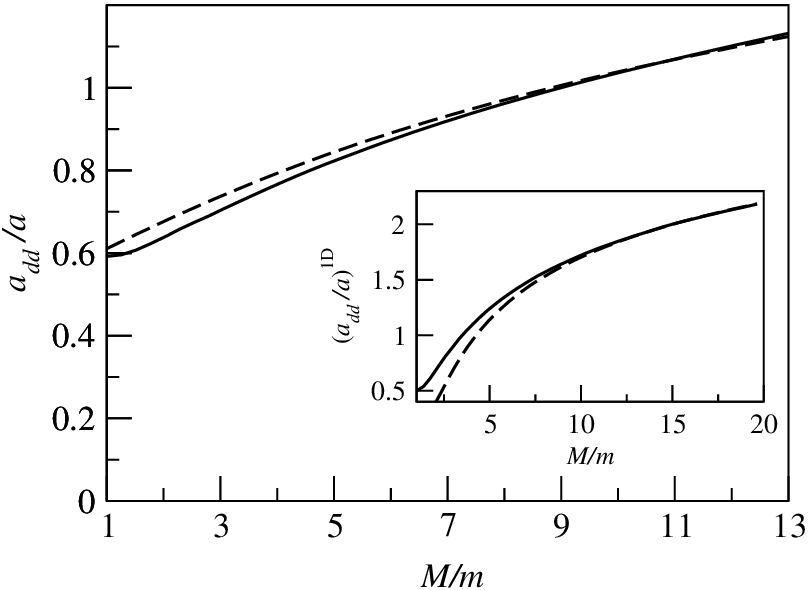}
\caption{
Dimer-dimer $s$-wave scattering length obtained by the exact four-body method (solid line) and by HBO (dashed line). The inset shows the results of a similar calculation in 1D.
\label{benchmark}}
\end{figure}

In certain aspects the accuracy of the HBO method can be limited by the accuracy of the BO approximation for the three-body problem. 
For example, the BO result for $\lambda_1$ in the case of Cs-Cs-Li three-body system is 5.3, whereas the exact value of this multiplicative factor is close to 4.9. 
Accordingly, the results presented in Fig.~\ref{RbCsLi} (and also in Fig.~\ref{YbLi}) can slightly deviate from the actual numbers. However, we believe that the HBO approach 
reproduces reasonably well the shape of the inelastic rate dependence on $a$ even for these not very large mass ratios and, more importantly, it captures the qualitative 
physics behind the processes of trimer formation and relaxation into deep bound states.

Finally, the wavefunction of a four-body system, in which one of the three-body subsystems exhibits Efimov physics, has many nodes (infinite in the zero-range approximation) 
in both real and momentum space. At present, as far as we know, there is no reliable method that can handle the very large configuration space arising in such problems. 
The HBO approach allows one to use a significantly reduced configuration space and is likely to be the only approach that is capable of solving such four-body problems 
for large mass ratios. 

\section{Concluding remarks}\label{VII}

Concluding the paper, we would like to emphasize several issues important for future studies. First of all, mixtures of light fermionic atoms with heavy fermions or bosons
are promising for the observation of the Efimov effect. The search for Efimov trimer states was in the agenda of physicists for more than 30 years, mostly in nuclear and
atomic physics \cite{efimov70,jensen04,braaten07}. The main reason is a remarkable discrete scaling symmetry of the system: dimensionless three-body observables are invariant under
the 
transformation $a\rightarrow\lambda^n a$, where $n$ is any integer and the value of $\lambda$ is fixed by Eq.~(\ref{lambda}). For example, invariant are the ratios of binding
energies to
$\epsilon_0$ or the ratio of the atom-dimer scattering length to $a$. However, the evidence for 
Efimov states was obtained only recently in Innsbruck experiments with cesium atoms through the measurements of the rate of three-body recombination as a function of $a$
\cite{kraemer06,naegerl06}. 
An important aspect is that the factor $\lambda$ for a three-body system of one light and two heavy atoms can be relatively small, for example $\lambda\approx 5$ for the 
$^{133}$Cs-$^{133}$Cs-$^6$Li system instead 
of $22.7$ for the Cs$_3$ trimer. This can allow the observation of several oscillations of the log-periodic dependence of three-body observables in the same window for the values
of the
two-body scattering length $a$.

We should mention here that a direct observation of bound Efimov trimers can be rather difficult because of their short intrinsic lifetime. Indeed, considering a trimer with the
energy 
$|\epsilon_{\nu}| \approx 10 \mu$K and assuming an optimistically small $\eta_*\sim 10^{-2}$, we obtain the trimer lifetime of the order of 100 $\mu$s. Therefore, 
one should invent indirect ways of identifying the existence of Efimov trimers. The key idea that we bring in with this paper is that the observation of the Efimov effect can be
made 
in a gas of dimers. From our results for the dimer-dimer inelastic collisional rate
(trimer formation) in Figs.~\ref{YbLi} and \ref{RbCsLi} one can extract information on the binding energies of trimer states, their lifetimes, and also on the value of the
three-body
parameter. The dependence $\alpha(a)$ immediately gives the parameter $\eta_*$ and, hence, the trimer lifetime $\tau$. The positions of the minima of the 
curve $\alpha(a)$ indicate the values of the two-body scattering length $a$ at which one has a trimer state with the binding energy $2\epsilon_0$.

One can think of obtaining long-lived bound light-heavy-heavy trimer 
states in an optical lattice acting on heavy atoms and increasing their effective mass \cite{PAPSS}. The heavy atoms in these states will be localized in different lattice sites
and 
the light atom will be delocalized between them. For example, one can consider the $^{40}$K-$^{40}$K-$^6$Li system which does not exhibit the Efimov effect in free space and will
show
the presence of Efimov trimers in an optical lattice under an increase of the K mass by more than a factor of 2. 
The intrinsic lifetime of such trimers will be long as their decay into a $^6$Li-$^{40}$K (or $^{40}$K-$^{40}$K) deeply bound molecule and a free atom requires two $^{40}$K atoms
to be on 
the same lattice site, which is strongly suppressed by the Pauli principle.

\section*{ACKNOWLEDGEMENTS}

We acknowledge hospitality and support of Institut Henri Poincar\'e 
during the workshop ``Quantum Gases'' where part of this work was done.  
The work was also supported by the IFRAF Institute, by
ANR grants Nos. 05-BLAN-0205 and 06-NANO-014, by the Russian Foundation for Fundamental Research, by the QUDEDIS program of ESF, and by the Dutch Foundation FOM. B.M.~and
S.J.J.M.F.K.~acknowledge support from the Netherlands Organization for Scientific Research (NWO). LPTMS is a mixed research unit No. 8626 of CNRS and Universit\'e Paris Sud.  

\section*{APPENDIX}

Explicit expressions for the operators $\hat L$ and $\hat L'$ in the case of $s$-wave scattering [$f({\bf R})\equiv f_0(R)$] and $p$-wave scattering 
[$f({\bf R})\equiv f_1(R)\cos\theta_{\bf R}$] are obtained by performing the angular integration in Eqs.~(\ref{Lprime}) and (\ref{L}). The calculation is straightforward 
and here we present only the resulting expressions. 

In the case of $s$-wave scattering we get:
\begin{widetext}
\begin{eqnarray}
\displaystyle
\hat{L} f_0(R)& &= \frac{q\sin^2 \theta\cos\theta}{2\pi^2}P\int_0^{\infty}{\mathrm d}R'\frac{R'}{R}\left\{ \left[ \frac{K_1 \left(q |R-R'|/\sin\theta \right)}
{|R-R'|} -(R'\rightleftarrows -R')\right] (f_0(R)-f_0(R')) \right. \nonumber \\
&&\left. +\frac{1}{\cos{2\theta}} \left[ \frac{K_1 \left(q\sqrt{R^2+R'^2 -2 R R' \cos{2\theta}}/\sin\theta \right)}{\sqrt{R^2+R'^2 -2 R R' \cos{2\theta}}} -
(R'\rightleftarrows -R') \right] f_0(R') \right\}.
\end{eqnarray}
Here $q=\sqrt{M(\epsilon_0-E)}$ and in each of the square brackets the second expression is the same as the first one with $R'$ replaced by $-R'$.

The action of the operator $\hat{L'}$ on a spherically symmetric function reads:
\begin{equation} \label{solpartialswave}
\hat{L'} f_0(R) =\sum_{l=1,3,...} \frac{2l+1}{2\pi} \int_0^{\infty} \sum_{n}[\chi_{n,l}(R) \chi_{n,l}^{\ast}(R') K^{(l)}_{\kappa_{n,l}}(R,R')-\chi_{n,l}^0(R) \chi_{n,l}^{0\ast}(R')
K^{(l)}_{\kappa_{n,l}^0}(R,R')]
f_0(R')R'^2{\mathrm d}R',
\end{equation}
where $\chi_{n,l}(R)$ is the radial part of the eigenfunction of Eq.~(\ref{BO3body}) with orbital angular momentum $l$ and radial quantum number $n$. 
The quantities $\chi_{n,l}^0(R)$ and $\kappa_{n,l}^0$ correspond to solutions of Eq.~(\ref{BO3body}) in which $\epsilon_{+}(R)$ is substituted by $\epsilon_0$. The kernel
$K^{(l)}_{\kappa}$ is defined as 
\begin{equation} \label{greenpartial}
K^{(l)}_{\kappa}(R_1,R_2) = 2 I_{l+1/2}( \kappa R_</2) K_{l+1/2}( \kappa R_>/2)/\sqrt{R_1 R_2},
\end{equation}
where the functions $I_{\mu}(z)$ and $K_{\mu}(z)$ are the modified Bessel functions of the first and second kind, and $R_<$ ($R_>$) is the smallest (largest) of the 
coordinates $R_{1,2}$. For $\epsilon_{\nu}\equiv \epsilon_{n,l}<E$, i.e. for negative imaginary $\kappa_{n,l}$ given by Eq.~(\ref{kappa}), we use the rules 
$K(-iz)=(\pi i^{\mu+1}/2)H_\mu (z)$ and $I_\mu (-iz)=J_\mu (z)$.

In the $p$-wave case we define the radial operator $\hat{L}^{l=1}$ by the relation $\hat L[f_1(R)\cos\theta_{\bf R}]=\cos\theta_{\bf R}\hat{L}^{l=1} f_1(R)$. The explicit form is
given by
\begin{eqnarray}
\displaystyle
\hat{L}^{l=1} f_1(R) \!\!\!\!\! &&= \frac{q\sin^2\theta\cos\theta}{2\pi^2}P\int_0^{\infty}{\mathrm d}R'\frac{R'}{R}\left\{ \left[ \frac{K_1 (q |R-R'|/\sin\theta)}{|R-R'|} -
(R'\rightleftarrows -R') \right] 
f_1(R) \right. \nonumber \\
&&+\sin{\theta}\left[ \left(\frac{K_0 ( q |R-R'|/\sin\theta )}{qRR'}+\frac{K_0 ( q \sqrt{R^2+R'^2-2RR'\cos{2\theta}}/\sin\theta)}{\cos^2 2\theta qRR'}\right) +
(R'\rightleftarrows -R') \right] 
f_1(R')\nonumber \\ 
&&\left.-\left[ \left(\frac{K_1 (q |R-R'|/\sin\theta)}{|R-R'|}+\frac{K_1 (q \sqrt{R^2+R'^2-2RR'\cos{2\theta}}/\sin\theta)}{\cos 2\theta \sqrt{R^2+R'^2-2RR'\cos{2\theta}}} \right)+
(R'\rightleftarrows -R') 
\right]f_1(R') \right\},
\end{eqnarray}
and, accordingly, for the radial part of the operator $\hat {L'}$ we have
\begin{equation} \label{solpartialpwave}
(\hat{L'})^{l=1} f_1(R) =\frac{1}{2\pi}\sum_{l=0,2,...} \int_0^{\infty} \sum_{n}[\chi_{n,l}(R) \chi_{n,l}^{\ast}(R') \bar{K}^{(l)}_{\kappa_{n,l}}(R,R')-\chi_{n,l}^0(R) 
\chi_{n,l}^{0\ast}(R') \bar{K}^{(l)}_{\kappa_{n,l}^0}(R,R')]f_1(R')R'^2{\mathrm d}R',
\end{equation}
where the kernel $\bar{K}^{(l)}$ is given in terms of $K^{(l)}$ from Eq.~(\ref{greenpartial}):
\begin{equation}
\bar{K}^{(l)}_{\kappa_{n,l}} \left(R,R'\right) = (l+1) K^{(l+1)}_{\kappa_{n,l}}(R,R') +l K^{(l-1)}_{\kappa_{n,l}}(R,R').
\end{equation}
\end{widetext}

\end{document}